# An Innovative Imputation and Classification Approach for Accurate Disease Prediction

Yelipe UshaRani

Department of Information Technology
VNR VJIET
Hyderabad, INDIA

Dr.P.Sammulal

Dept.of Computer Science and Engineering
JNT University
Karimnagar, INDIA

*Abstract*—**Imputation of missing attribute values in medical datasets for extracting hidden knowledge from medical datasets is an interesting research topic of interest which is very challenging. One cannot eliminate missing values in medical records. The reason may be because some tests may not been conducted as they are cost effective, values missed when conducting clinical trials, values may not have been recorded to name some of the reasons. Data mining researchers have been proposing various approaches to find and impute missing values to increase classification accuracies so that disease may be predicted accurately. In this paper, we propose a novel imputation approach for imputation of missing values and performing classification after fixing missing values. The approach is based on clustering concept and aims at dimensionality reduction of the records. The case study discussed shows that missing values can be fixed and imputed efficiently by achieving dimensionality reduction. The importance of proposed approach for classification is visible in the case study which assigns single class label in contrary to multi-label assignment if dimensionality reduction is not performed.**

*Keywords— imputation; missing values; prediction; nearest neighbor, cluster, medical records, dimensionality reduction*

## I. INTRODUCTION

Medical records preprocessing is an important step which cannot be avoided in most of the situations and when handling medical datasets. The attributes present in medical records may be of different data types. Also, the values of attributes have certain domain which requires proper knowledge from medical domain to handle them.

This is because of this diverse nature of medical records, handling medical records is quite challenging for data miners and researchers. The various preprocessing techniques for medical records include fixing outliers in medical data, estimation and imputing missing values, normalizing medical attributes, handling inconsistent medical data, applying smoothing techniques to attributes values of medical records to specify some of them.

Data Quality depends on Data Preprocessing techniques. An efficient preprocessing of medical records may increase the data quality of medical records. In this context, data preprocessing techniques have achieved significant importance from medical data analysts and data miners. Incorrect and improper data values may mislead the prediction and classification results, there by resulting in false classification results and thus leading to improper medical treatment which is a very dangerous potential hazard. This research mainly aims at handling missing attribute values present in medical records of a dataset. The attributes may be numeric, categorical etc. The present method can handle all the attribute types without the need to devise a different method to handle different attribute types. This is first importance of our approach. We outline research objective and problem specification in the succeding lines of this paper and then discuss importance of our approach.

### A. Research Objective

We have the following research objectives in this research towards finding missing values

- Obviously our first and foremost objective is to impute missing values.

- Aim at dimensionality reduction process of medical records.

- Classify new medical records using the same approach used to find missing values.

- Cluster medical records to place similar records in to one group.

### B. Problem Specification

Given a dataset of medical records with and without missing values, the research objective is to fix set of all missing values in the medical records by using a novel efficient Imputation approach based on clustering normal medical records, so as to estimate missing values in medical records with missing values.

### C. Importance of Present Approach

The importance of the present approach which we wish to propose has the following advantages

- The method may be used to find missing attribute values from medical records

- The same approach for finding missing values may be used to classify medical records

- The disease prediction may be achieved using the proposed approach without the need to adopt a separate procedure

- Handles all attribute types

- Preserves attribute information

- May be applied for datasets with and without class labels which is uniqueness of the current approach.

## II. RELATED WORKS

Most of the research works carried in the literature argues that the presence of missing values of medical attributes makes the extra overhead may be in prediction and classification or when performing dataset analysis. In contrast to this the researchers Zhang, S et.al in their work [1] discuss and argue that missing values are useful in cost sensitive environments [17-18]. This is because some of the attributes values incur high cost to fill those values by carrying experiments. In such cases, it would be cost effective to skip such tests and values associated with those medical tests. Handling Missing values in medical datasets is quite challenging and also requires use of statistical approaches [15, 16] to estimate the same. In [2], missing values are found by using clustering approach where the missing value is filled with the value of attribute of nearest clusters. The concept of support vector regression and clustering is applied to find missing values in the work of authors [3]. In [4], phylogeny problems occurring because of missing values is discussed. An approach to handle medical datasets consisting mixed attribute types is handled in [5]. Some of the research contributions in missing values include [6-21].

## III. RESEARCH ISSUES IN MINING MEDICAL DATA

### A. Handling Medical Datasets

The research should first start with the studying the benchmark datasets. Sometimes there may be a need to start collecting data from scratch if we are working over a problem in particular domain. However, when working with medical datasets, we need to remember that the dataset is multi-variate.

### B. Handling Missing Values in Medical Datasets

The medical datasets are not free from missing values. Obviously there is no free lunch. We must make sure to handle the missing values suitably and accurately. A simple approach would be to discard the whole record which essentially contains the   missing value of an attribute. Some significant novel approaches include [6-8, 12-14].

### C. Choice of Prediction and Classifications algorithms

The underlying dataset is the deciding factor for choice of the algorithm. A single classification algorithm is not suitable for every dataset. Recent works include [18-19].

### D. Finding Nearest Medical Record and Identifying the Class Label

The heart of any classification or clustering algorithm is the distance measure used to estimate the record distance between any two records. Since classification involves training and testing phases, training dataset must include all possible combinations which forms a knowledge database using which class label is estimated accurately. Finding nearest records may be performed through using KNN-classifiers or using any other classifiers. Classification may has a curse of dimensionality. Hence, dimensionality reduction must be suitably addressed. However, this can make situation complex and also inaccurate sometimes, if important attribute information or attributes are missed or discarded [9-11].

### E. Deciding on Medical Attributes

The attributes of the medical dataset are also the prime concern in prediction and classification. This is because the attributes are multi variants [11]. Coming up with the deciding attributes for heart disease prediction which can make significant impact on the classification accuracy and prediction of the disease symptom is also one of the important tasks. In short, it is required to perform a thorough literature survey, fix the attributes which must be considered and which may be discarded.

### F. Removal of Noise

After deciding the number of attributes, we may have one or more attributes which may be not important and hence may be discarded without any loss. Every effort must be made in this direction, so that the attributes which are of least importance and removal of the attributes does not make any significant affect may be eliminated.

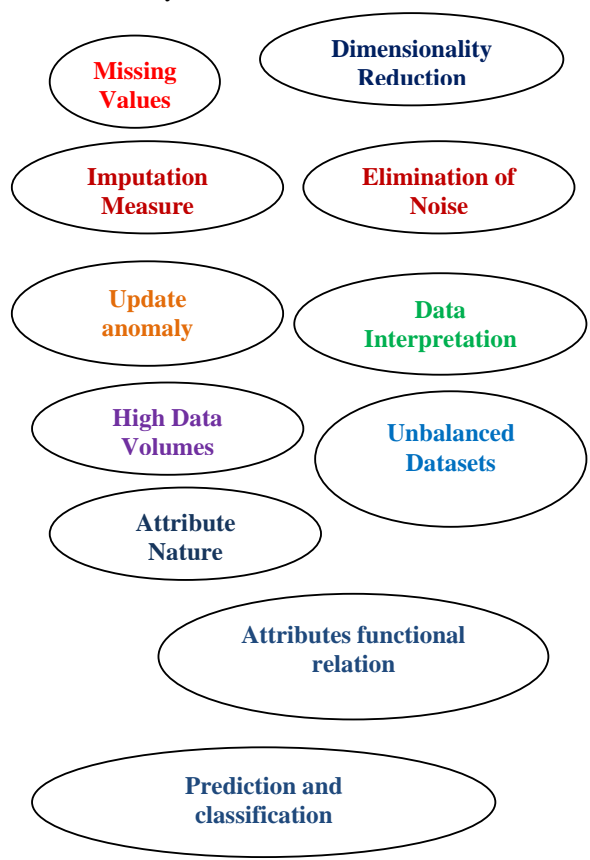

Fig. 1    Research Problems when handling Medical Datasets

## IV.  IMPUTATION FRAME WORK

In this section, we discuss framework to impute missing values as shown in Fig.2 and Fig.3

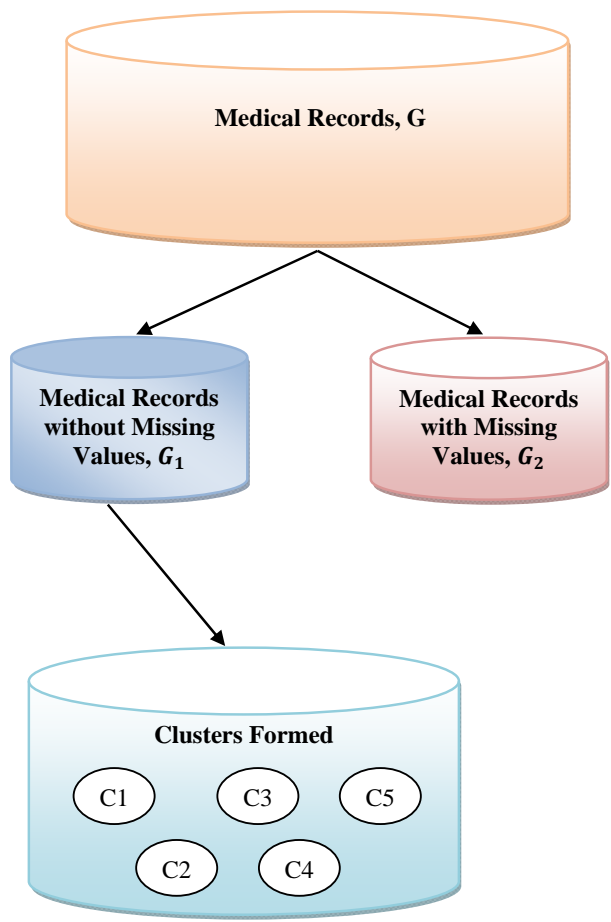

Fig. 2  Generating Clusters from medical records

The framework for missing value Imputation consists of following steps. The approach for missing values is based on the concept of clustering medical records without missing values. This is because, all similar medical records shall come into one cluster and hence imputation performed shall be more accurate. This approach of finding missing values has not been carried out earlier in the literature. We present analytical framework with a case study in this paper. This research was motivated form the work by researchers for intrusion detection published in 2015 [20].

### A.  Generating Clusters from Group $G_1$

- This step involves finding the number of class labels and generating number of clusters equal to number of class labels

- The clusters may be generated using k-means algorithm by specifying value of k to be number of class labels.

Alternately, we may apply any clustering algorithm which can generate k clusters

### B.  Computing distance of normal records to Cluster Centers

- Obtain mean of each cluster. This shall be the cluster center

- Obtain distance of each medical record to each cluster center.

- Sum all distances obtained

- The result is all medical records mapped to single value achieving dimensionality reduction.

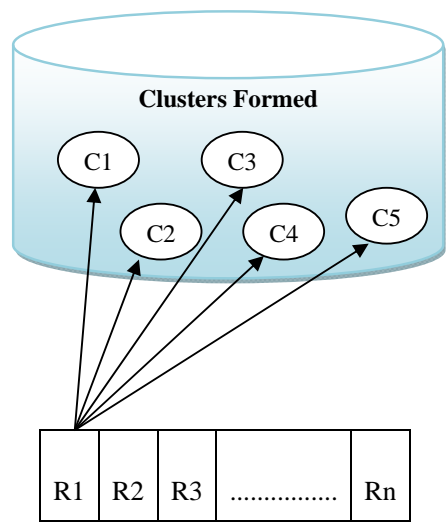

Fig. 3  Distance Computation from Record to Cluster Centers

### C.  Computing distance of missing records to Cluster Centers

- Obtain distance of each medical record having missing values to each cluster center by discarding those attributes with missing values.

- Sum all distances obtained

- The result is all medical records mapped to single value achieving dimensionality reduction.

- Method preserves information of attributes

### D.  Find Nearest Record to Impute Missing Values

Consider each missing record in group, $G_2$ one by one. Find the distance of this record to all the records in group $G_1$. The record to which the distance is minimal, shall be the nearest neighbor. Perform imputation of the missing attribute value by considering the corresponding attribute value of nearest record in that class. The frequency may also be considered for imputation incase, we have more than one nearest neighbors.

## V. PROPOSED IMPUTATION ALGORITHM

### A. Proposed Algorithm

**Input**: *Medical Records with Missing Values*

**Output**: *Imputation of Missing Values*

**Notations adopted**:

$R_i$ — $i^{th}$ medical record

$R_i(A_K)$ — $k^{th}$ attribute value of $i^{th}$ medical record

$G_c$ — $c^{th}$ group

$i, k$ — index of medical records and attributes

$\emptyset$ — misisng record or Empty record value

$c$ — number of decision classes in medical dataset

$D_d$ — $d^{th}$ decision class

$m$ — total number of medical records

$n$ — number of attributes in each record

$\mu_d$ — cluster center of $d^{th}$ cluster

$\mu_{dn}$ — mean value of $n^{th}$ attribute

$h$ — number of records in group, $G_2$

$z$ — number of records in group, $G_1$ equal to $(m - h)$

### Step-1: Read Medical Dataset

Read the medical dataset consisting of medical records. Find records with and without missing values. Classify records in to two groups, say $G_1$ and $G_2$. The first group, $G_1$ is set of all medical records with no missing values given by Eq.1. The second group, $G_2$ is set of all medical records having missing values given by Eq.2.

$$G_1 = U \{ R_i \mid R_i(A_K) \neq \emptyset, \forall i, k \} \quad (1)$$

$$G_2 = U \{ R_i \mid R_i(A_K) = \emptyset / \exists i, k \} \quad (2)$$

Where $i \in (1, m - h)$ and $k \in (1, n)$. We may consider group, $G_1$ as training set of medical records while group, $G_2$ is considered as testing set in this case.

### Step-2: Cluster Medical Records with No Missing values

Let, $g = |D_d|$, be the number of decision classes. Determine the maximum number of decision classes available in the medical dataset being considered. Cluster the medical records in group, $G_1$ to a number of clusters equal to g. i.e $|D_d|$.

This may be achieved using K-means clustering algorithm. This is because K-means algorithm requires the number of required clusters to be specified well ahead before clustering process is carried out. The output of step-2 is a set of clusters. i.e Number of output clusters is equal to 'g'.

This is shown in fig.4 and fig.5 where a set of medical records represented by $G_1$ are clustered in to 'd' clusters.

### Step-3: Obtain Cluster Center for each Cluster formed

This step involves finding the cluster center for each cluster which is generated using the k-means clustering algorithm. We can obtain the cluster center by finding the mean of each attribute from attribute set, $A_K$ of medical attributes.

Let Cluster- $C_d$ denotes $d^{th}$ cluster having the records $R_1$, $R_6$, $R_8$ and $R_9$ with single attribute. Then the cluster center is given by Eq.3 as

$$\mu_d = \frac{R_1(A_1) + R_6(A_1) + R_8(A_1) + R_9(A_1)}{4} \quad (3)$$

In general the cluster center of $g^{th}$ cluster may be obtained using the generalized equation, Eq.4 given below

$$\mu_g = U_k \left[ \frac{\{\sum R_l^k \mid l \in \{1, q\} \text{ for each } k \in \{1, n\} \}}{|l|} \right] \quad (4)$$

$\mu_g$ is hence a sequence of 'n' values indicating cluster center over 'n' attributes. The notation, $U_k$ is used to denote set of all values each separated by a symbol comma. The cluster center may hence be formally represented using the representation given by Eq.5

$$\mu_g = < \mu_{g1}, \mu_{g2}, \mu_{g3}, \mu_{g4}, \ldots \ldots \mu_{gn} > \quad (5)$$

Here 'n' indicates total number of attributes in each medical record and |g| indicates number of clusters.

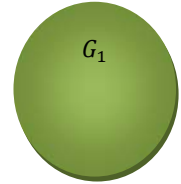

Fig 4.    $G_1$ Before Clustering

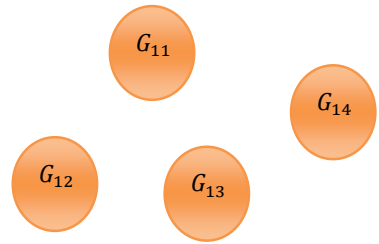

Fig. 5   Before and After Clustering

### Step-4: Compute distance between each $R_i$ and each $\mu_d$

Find distance from each medical record, $R_i$ in group, $G_1$ to each of the cluster centers, $\mu_d$ obtained in step-3. This can be achieved through finding the Euclidean distance between each medical record of 'n' attributes and cluster center of each cluster defined over 'n' attributes. These cluster distances computed are summed to obtain a single distance value. This distance is called Type-1 distance value given by Eq. 6.

$$\text{Dist}^d(R_i, \mu_d) =$$

$$\sqrt{(R_{i1} - \mu_{d1})^2 + (R_{i2} - \mu_{d2})^2 + \cdots (R_{in} - \mu_{dn})^2} \quad (6)$$

$$\forall i \in (1, n), \forall d$$

At the end of Step-4 we have distance value from each record , $R_i$ to each cluster center denoted by $\mu_d$.

**Step-5: Transform multi-dimensional medical record to a single dimension numeric value by using mapping function**

$Map(R_i)$ is a mapping function which maps the medical record, $R_i$ to a single distance value. To determine mapping function value of a record we use the equation, Eq.7

This can be obtained by adding all distances obtained in Step-4

$$Map(R_i) = \sum_{d=1}^{d=|g|} \sum \text{Dist}^d(R_i, \mu_d) \qquad \forall i \in (1, n-h) \quad (7)$$

Where |g| is number of clusters formed, (n-h) indicates number of records in group, $G_1$.

At the end of step-5, we have each medical record, $R_i$ mapped to a single dimension distance value. In other words, the medical record of 'n' dimensions is reduced to single dimension achieving dimensionality reduction.

**Step-6: Compute distance between each $R_j$ in group, $G_2$ and each $\mu_d$ of clusters formed**

Obtain distance value of missing records to these cluster centers by discarding the attributes with missing values. Find distance from each medical record, $R_j$ in group, $G_2$ to each of the cluster centers, $\mu_d$ obtained in step-3.

This can be achieved through finding the Euclidean distance between each medical record of 'n' attributes and cluster center of each cluster defined over 'n' attributes. These cluster distances computed are summed to obtain a single distance value. This distance is called Type-2 distance value given by Eq.8.

$$\text{Dist}^2(R_{in}, \mu_{dn})$$

$$= \sqrt{\sum (R_{i1} - \mu_{d1})^2 \ldots \ldots \ldots} \forall R_{in} \text{ where } n \neq y \quad (8)$$

discarding the $y^{th}$ missing attribute value.

**Step-7: Transform multi-dimensional medical record with missing values to a single dimension by using mapping function**

$Map^r(R_j)$ is a mapping function which maps each medical record, $R_j$ consisting to a single distance value. To determine mapping function value of a record $R_j$, we use the equation, Eq.9

$$Map^r(R_j) = \sum_{d=1}^{d=|g|} \sum \text{Dist}^d(R_j, \mu_d) \qquad \forall j \in (1, h) \quad (9)$$

Where |g| is number of clusters formed and $j \in (1, h)$

At the end of step-7, we have each medical record, $R_j$ mapped to a single dimension distance value. In other words, the medical record of 'n' dimensions is reduced to single dimension achieving dimensionality reduction.

**Step-8: Obtain difference between distances obtained in step-6 and step-8**

For each missing record, $R_j$ in $G_2$, obtain difference between mapping functions of each record, $R_i$ in group, $G_1$ and missing record, $R_j$ in group, $G_2$. Call this value as $d_{ij}$

**Step-9: Find nearest record**

The medical record $R_j$ is most similar to the medical record, $R_i$ whose corresponding $d_{ij}$ is most minimum as given by Eq.10.

$$d_{ij} = Min_i \{Map(R_i) - Map^r(R_j)\} \quad (10)$$

**Step-10: Fix Missing values and Impute Missing Values**

The medical record $R_j$ is most similar to the medical record, $R_i$ whose corresponding $d_{ij}$ is most minimum. In this case, impute the missing attribute value of record, $R_j$ denoted by $R_{jr}$ by the attribute value, $R_{ir}$ of medical record denoted as $R_i$.

Incase more than one record with same minimum value is obtained then, fill the missing value of the attribute with the attribute value whose frequency is maximum from the same decision class. Alternately, we may fix the mean of the values also from the corresponding decision class attribute values.

## VI. CASE STUDY

In this Section-VI, we discuss case study to find missing attributes values of medical records by using the proposed approach. For this, we consider a sample dataset consisting sample values.

Consider Table. I, shown below consisting of sample dataset of medical records having categorical and numerical values. Table. II shows medical records without missing values after normalizing sample dataset. Table.III denotes records with and without missing values. Table IV denotes all records without missing values and Table. V shows records with missing attribute values.

Table.VI depicts clusters generated from group $G_1$, which consists medical records with no missing values after applying k-means algorithm. There are two clusters generated $C_1$ and $C_2$.

$C_1$ contains set of all medical records {$R_1, R_4, R_6, R_9$} and $C_2$ contains set of all medical records {$R_2, R_7, R_8$}. Table.VII gives the distances of records in group, $G_1$ to cluster center of the first cluster. Similarly, Table.VIII gives the distances of records in group, $G_2$ to cluster center of the second cluster.

Table.IX depicts computation values of mapping function of records of group, $G_1$. The mapping function $Map(R_i)$ is mapping distance of $i^{th}$ record, which is sum of all distances from record, $R_i$ to each of those cluster centers generated from application of clustering algorithm.

Table. X gives the distances of medical records in group, $G_2$ to each of the cluster centers.

Table. XI depicts computation values of mapping function of medical records containing missing values of group, $G_2$. The mapping function $Map^r(R_j)$ is mapping distance of $j^{th}$ record, which is sum of all distances from record, $R_j$ to each of those cluster centers generated from application of clustering algorithm. The distance is computed considering those attributes which do not have missing values. i.e Attribute values are defined and recorded.

TABLE I.    DATASET OF MEDICAL RECORDS

| Record | A1 | A2 | A3 | A4 | Decision Class |
|--------|-----|-----|----------|-----|----------------|
| R1 | $c_{11}$ | 5 | $d_{31}$ | 10 | CLASS-1 |
| R2 | $c_{13}$ | 7 | $d_{31}$ | 5 | CLASS-1 |
| R3 | $c_{11}$ | 7 | $d_{32}$ | 7 | CLASS-1 |
| R4 | $c_{12}$ | 5 | $d_{31}$ | 10 | CLASS-1 |
| R5 | $c_{13}$ | 3 | $d_{32}$ | 7 | CLASS-2 |
| R6 | $c_{12}$ | 9 | $d_{31}$ | 10 | CLASS-2 |
| R7 | $c_{11}$ | 5 | $d_{32}$ | 3 | CLASS-2 |
| R8 | $c_{13}$ | 6 | $d_{32}$ | 7 | CLASS-2 |
| R9 | $c_{12}$ | 6 | $d_{32}$ | 10 | CLASS-2 |

TABLE II.    NORMALIZED DATASET OF RECORDS

| Record | A1 | A2 | A3 | A4 | Decision Class |
|--------|-----|-----|-----|-----|----------------|
| R1 | 1 | 5 | 1 | 10 | CLASS-1 |
| R2 | 3 | 7 | 1 | 5 | CLASS-1 |
| R3 | 1 | 7 | 2 | 7 | CLASS-1 |
| R4 | 2 | 5 | 1 | 10 | CLASS-1 |
| R5 | 3 | 3 | 2 | 7 | CLASS-2 |
| R6 | 2 | 9 | 1 | 10 | CLASS-2 |
| R7 | 1 | 5 | 2 | 3 | CLASS-2 |
| R8 | 3 | 6 | 2 | 7 | CLASS-2 |
| R9 | 2 | 6 | 2 | 10 | CLASS-2 |

TABLE III.    RECORDS WITH AND WITHOUT MISSING VALUES

| Record | A1 | A2 | A3 | A4 | Decision Class |
|--------|-----|-----|------|-----|----------------|
| R1 | 1 | 5 | 1 | 10 | CLASS-1 |
| R2 | 3 | 7 | 1 | 5 | CLASS-1 |
| R3 | 1 | 7 | NaN | 7 | CLASS-1 |
| R4 | 2 | 5 | 1 | 10 | CLASS-1 |
| R5 | 3 | 3 | 2 | NaN | CLASS-2 |
| R6 | 2 | 9 | 1 | 10 | CLASS-2 |
| R7 | 1 | 5 | 2 | 3 | CLASS-2 |
| R8 | 3 | 6 | 2 | 7 | CLASS-2 |
| R9 | 2 | 6 | 2 | 10 | CLASS-2 |

TABLE IV.    RECORDS WITH OUT MISSING VALUES

| Record | A1 | A2 | A3 | A4 | Decision Class |
|--------|-----|-----|-----|-----|----------------|
| R1 | 1 | 5 | 1 | 10 | CLASS-1 |
| R2 | 3 | 7 | 1 | 5 | CLASS-1 |
| R4 | 2 | 5 | 1 | 10 | CLASS-1 |
| R6 | 2 | 9 | 1 | 10 | CLASS-2 |
| R7 | 1 | 5 | 2 | 3 | CLASS-2 |
| R8 | 3 | 6 | 2 | 7 | CLASS-2 |
| R9 | 2 | 6 | 2 | 10 | CLASS-2 |

TABLE V.    RECORDS WITH MISSING VALUES

| Record | A1 | A2 | A3 | A4 | A5 | Decision Class |
|--------|-----|-----|-----|-----|-----|----------------|
| R3 | 1 | 7 | ? | 7 | 1 | CLASS-1 |
| R5 | 3 | 3 | 2 | ? | 3 | CLASS-2 |

TABLE VI.    CLUSTERS GENERATED USING K-MEANS

| Clusters | Medical Records with out missing values |
|----------|----------------------------------------|
| C1 | R1,R4,R6,R9 |
| C2 | R2,R7,R8 |

TABLE VII.    DISTANCE OF RECORDS TO CLUSTER-1

| Record | Distance to Cluster-1 |
|--------|----------------------|
| R1 | 5.312459 |
| R2 | 1.374369 |
| R4 | 5.153208 |
| R6 | 5.878397 |
| R7 | 2.624669 |
| R8 | 2.134375 |
| R9 | 5.022173 |

TABLE VIII.    DISTANCE OF RECORDS TO CLUSTER-2

| Record | Distance to Cluster-2 |
|--------|----------------------|
| R1 | 1.47902 |
| R2 | 5.214163 |
| R4 | 1.299038 |
| R6 | 2.772634 |
| R7 | 7.189402 |
| R8 | 3.344772 |
| R9 | 0.829156 |

TABLE IX.    MAPPING DISTANCE OF MEDICAL RECORDS WITHOUT MISSING VALUES

| Record | Mapping Distance |
|--------|------------------|
| R1 | 6.791479 |
| R2 | 6.588532 |
| R4 | 6.452246 |
| R6 | 8.651031 |
| R7 | 9.814071 |
| R8 | 5.479147 |
| R9 | 5.851329 |

TABLE X.    MISSING VALUE RECORD DISTANCES TO CLUSTERS

| Record | Distance to Cluster-1 | Distance to Cluster-2 |
|--------|----------------------|----------------------|
| R3 | 2.603417 | 3.181981 |
| R5 | 3.091206 | 3.561952 |

TABLE XI.    MAPPING DISTANCE OF MEDICAL RECORDS WITH MISSING VALUES

| Record | Mapping Distance |
|--------|------------------|
| R3 | 6.791479 |
| R5 | 6.588532 |

TABLE XII.  DISTANCE OF MEDICAL RECORDS R3 WITH OTHER RECORDS

| Record | Distance  with R3 |
|--------|-------------------|
| R1 | 0 |
| R2 | -0.20295 |
| R4 | -0.33923 |
| R6 | 1.859552 |
| R7 | 3.022592 |
| R8 | -1.31233 |
| R9 | -0.94015 |

Table.XII shows distance of record, $R_3$ to records $R_1$, $R_2$, $R_4$, $R_6$, $R_7$, $R_8$, $R_9$ . The record $R_3$ is nearest to medical record $R_8$. The Table. XIII shows nearest medical record $R_8$ for $R_3$ which is ideal record to carry imputation. The attribute value to be imputed is 2. i.e the categorical attribute value $d_{32}$. This is because the attribute value, $d_{32}$ was mapped to numerical value 2.

TABLE XIII.  NEASREST MEDICAL RECORD FOR RECORD R8

| Record | A1 | A2 | A3 | A4 | Decision Class |
|--------|----|----|----|----|----------------|
| R8 | 3 | 6 | 2 | 7 | CLASS-2 |

TABLE XIV.  DISTANCE OF MEDICAL RECORD R5 TO OTHER RECORDS

| Record | Distance  with R5 |
|--------|-------------------|
| R1 | 0.202947 |
| R2 | 0 |
| R4 | -0.13629 |
| R6 | 2.0625 |
| R7 | 3.225539 |
| R8 | -1.10939 |
| R9 | -0.7372 |

TABLE XV.  NEASREST MEDICAL RECORD FOR RECORD R5

| Record | A1 | A2 | A3 | A4 | Decision Class |
|--------|----|----|----|----|----------------|
| R8 | 3 | 6 | 2 | 7 | CLASS-2 |

Table.XIV shows distance of record, $R_5$ to records $R_1$, $R_2$, $R_4$, $R_6$, $R_7$, $R_8$, $R_9$ . The record $R_5$ is nearest to medical record $R_8$. The Table. XV shows nearest medical record $R_8$ for $R_5$ which is ideal record to carry imputation. The attribute value to be imputed is 7. i.e the numerical value.

Finally in this case study, we fill the missing values of medical records by imputing the missing attribute values. Since, attribute values after imputing, happen to be the same values which were present initially in Table.1 the correctness of the approach can be verified and validated. The proposed approach of finding imputation value is hence accurate and also efficient as it also aims at dimensionality reduction of medical records and then estimates missing values. In the process of dimensionality reduction we never miss any attribute values or attributes. This brings the accuracy in the present approach.

This approach may be extended to classify new medical record without class label to an appropriate class, if required by simply assigning class label of medical record to which the new record distance minimum. In this way, disease prediction or classification may be achieved.

## VII.  CLASSIFICATION OF NEW MEDICAL RECORDS

Consider the table of medical records with class labels as in Table.XVI with the parameter values same as Table. II, the last column is decision class, which predicts the disease level or stage. This table is free from missing values and hence is suitable for mining medical records.

TABLE XVI.  NORMALIZED MEDICAL RECORDS WITH CLASSES

| Record | P1 | P2 | P3 | P4 | Disease Class or Type |
|--------|----|----|----|----|-----------------------|
| R1 | 1 | 5 | 1 | 10 | Level-1 |
| R2 | 3 | 7 | 1 | 5 | Level-1 |
| R3 | 1 | 7 | 2 | 7 | Level-1 |
| R4 | 2 | 5 | 1 | 10 | Level-1 |
| R5 | 3 | 3 | 2 | 7 | Level-2 |
| R6 | 2 | 9 | 1 | 10 | Level-2 |
| R7 | 1 | 5 | 2 | 3 | Level-2 |
| R8 | 3 | 6 | 2 | 7 | Level-2 |
| R9 | 2 | 6 | 2 | 10 | Level-2 |

Assume that, we have an incoming medical record with the attribute values as R10 = [2, 5, 2, 9]. We can obtain Euclidean distances from record R10 to all the records R1 through R9. The class of the record is the class of medical record to which the Euclidean distance is minimum. Table. XVII gives distance of medical record R10 to all records.

TABLE XVII.  NORMALIZED MEDICAL RECORDS WITH CLASSES

| Record | Distance with R10 |
|--------|-------------------|
| R1 | 1.732051 |
| R2 | 4.690416 |
| R3 | 3 |
| R4 | 1.414214 |
| R5 | 3 |
| R6 | 4.242641 |
| R7 | 6.082763 |
| R8 | 2.44949 |
| R9 | 1.414214 |

Using this approach we get two class labels as the record is nearest to both records R4 and R8. But the classes are class-1 and classs-2 for R4 and R8 respectively. So we can't categorize the disease correctly or accurately. This is because; we did not perform dimensionality reduction. This is overcome if we extend the approach for fixing missing values to classification also. The only difference is that we continue to extend the procedure outlined in Section –IV, to all the records after fixing missing values (R1 to R9) and adopt the procedure for missing record to the new record but considering all attribute values. This is shown in computations below.

Table. XVIII shows clusters generated using k-means with K=2. Table.XIX and Table.XX shows distance of records R1 to R9 w.r.t cluster centers. Table XXI and Table XXIII shows mapping distance of R1 to R9 and Record R10 respectively.

TABLE XVIII.  CLUSTERS GENERATED USING K-MEANS

| Clusters | Medical Records with out missing values |
|---|---|
| C1 | R1,R4,R6,R9 |
| C2 | R2,R7,R8,R3,R5 |

Table.XXII gives distance value of new records R10 to clusters formed. Table. XXIV gives difference of mapping values of existing records and new record.

TABLE XIX.  DISTANCE OF RECORDS TO CLUSTER-1

| Record | Distance to Cluster-1 |
|---|---|
| R1 | 1.47902 |
| R2 | 5.214163 |
| R3 | 3.269174 |
| R4 | 1.299038 |
| R5 | 4.656984 |
| R6 | 2.772634 |
| R7 | 7.189402 |
| R8 | 3.344772 |
| R9 | 1.47902 |

TABLE XX.  DISTANCE OF RECORDS TO CLUSTER-2

| Record | Distance to Cluster-2 |
|---|---|
| R1 | 4.481071 |
| R2 | 1.969772 |
| R3 | 2.209072 |
| R4 | 4.322037 |
| R5 | 2.979933 |
| R6 | 5.46626 |
| R7 | 3.11127 |
| R8 | 1.509967 |
| R9 | 4.481071 |

TABLE XXI.  MAPPING DISTANCE OF MEDICAL RECORDS

| Record | Mapping Distance |
|---|---|
| R1 | 5.960091 |
| R2 | 7.183935 |
| R3 | 5.478246 |
| R4 | 5.621075 |
| R5 | 7.636917 |
| R6 | 8.238894 |
| R7 | 10.30067 |
| R8 | 4.854739 |
| R9 | 5.960091 |

TABLE XXII.  DISTANCES OF NEW MEDICAL RECORD TO CLUSTERS

| Record | Distance to Cluster-1 | Distance to Cluster-2 |
|---|---|---|
| R10 | 1.785357 | 3.628027 |

TABLE XXIII.  MAPPING DISTANCE OF R10

| Record | Mapping Distance |
|---|---|
| R10 | 5.053384 |

TABLE XXIV.  DIFFERENCE OF MAPPING DISTANCE OF NEW MEDICAL RECORD TO MAPPING DISTANCE OF EXISTING RECORDS

| Record | Distance with R3 |
|---|---|
| R1 | 0.906707 |
| R2 | 2.130551 |
| R3 | 0.424862 |
| R4 | 0.567691 |
| R5 | 2.583533 |
| R6 | 3.18551 |
| R7 | 5.247288 |
| R8 | -0.19865 |
| R9 | 0.906707 |

The class label of record R10 to be assigned is the class label of the record to which the distance is minimum in Table.XXIV. In this case, the distance of R10 is proved to be minimum w.r.t R8 as compared to other record distances. Hence the class label of the new record R10 is class label of record R8. i.e Class-2 or Level-2.

Hence, the category of disease level of person whose medical record values are defined by R10 is level-2 or class-2 or Type-2 disease.

## VIII.  ADVANTAGE OF PROPOSED METHOD

If we can see the result obtained with traditional approach without dimensionality reduction carried out, the category of disease is either Class-1 or Class-2. This is because of noise attribute values. We overcome such a disadvantage using the proposed method of dimensionality reduction and classification. Using this proposed method, we get a single class label for the new record. In our case, the class is identified as Class-2 or Level-2 using proposed method of classification. This is because of dimensionality reduction performed to single value without missing any attribute or neglecting any attribute value.

## IX.  DISCUSSIONS AND OUTCOMES

In this research, we address the first challenge of handling medical records in datasets. We discuss the approach for imputing missing attribute values of medical records. This is done by clustering medical records which were free from missing values. The records with missing values were separate from dataset. The multi-dimensional medical records are transformed to single dimension. In future, the objective is to see the possibility of other clustering procedures and new approaches to impute missing values. The present approach may be extended to perform classification and prediction without the need for adopting separate procedures to achieve the required objectives. This method is first of its kind which may be used to perform missing values imputation, classification, and disease prediction in a single stretch. A simple common sense shows the importance of the approach carried out and may be extended to any other domain of interest by researchers.

## X. CONCLUSIONS AND SCOPE FOR FUTURE RESEARCH

In the present research, we address the first challenge of handling missing values in medical datasets. We also address how the dimensionality reduction of medical datasets may be achieved in a simple approach. We come up with a new approach of finding missing values in datasets not addressed in the literature by aiming at a single dimension. The approach followed does not miss any attribute information while carrying out dimensionality reduction which is the importance of this approach. The proposed approach of imputing missing values in medical records is feasible for both categorical and numerical attributes as discussed in case study. However, suitable normalization techniques must be applied, if required for some datasets after extensive study of the datasets. In this paper, we also extend imputation approach also for prediction and classification of unknown medical records for predicting disease levels or symptoms through soft computing techniques. The approach overcomes ambiguity which is otherwise possible if dimensionality reduction is not carried properly.

## REFERENCES

[1] Zhang, S, Zhenxing Qin, Ling C.X, Sheng S, " "Missing is useful": missing values in cost-sensitive decision trees,", IEEE Transactions on Knowledge and Data Engineering, vol.17, no.12, pp.1689-1693, 2005.

[2] Zhang, C,Yongsong Qin, Xiaofeng Zhu, Jilian Zhang, and Zhang,S, "Clustering-based Missing Value Imputation for Data Preprocessing," in , 2006 IEEE International Conference on Industrial Informatics, pp.1081-1086, 2006.

[3] Wang, Ling, Fu Dongmei, Li Qing, Mu Zhichun, "Modelling method with missing values based on clustering and support vector regression," , Journal of Systems Engineering and Electronics , vol.21, no.1, pp.142-147, 2010.

[4] Kirkpatrick B, Stevens K, " Perfect Phylogeny Problems with Missing Values," IEEE/ACM Transactions on Computational Biology and Bioinformatics,Vol.11,No.5,pp.928-941,2014.

[5] Xiaofeng Zhu, Zhang S, Zhi Jin, Zili Zhang, and Zhuoming Xu, "Missing Value Estimation for Mixed-Attribute Data Sets", IEEE Transactions on Knowledge and Data Engineering, Vol.23, No.1, pp.110-121, 2011.

[6] Farhangfar A, Kurgan L.A, Pedrycz ,"A Novel Framework for Imputation of Missing Values in Databases," in Part A: Systems and Humans, IEEE Transactions on Systems, Man and Cybernetics, Vol.37, No.5,pp.692-709, 2007.

[7] Miew Keen Choong,Charbit M, Hong Yan, "Autoregressive-Model-Based Missing Value Estimation for DNA Microarray Time Series Data,",IEEE Transactions on Information Technology in Biomedicine,Vol.13, No.1,pp.131-137, 2009.

[8] Qiang Yang, Ling C, Xiaoyong Chai, and Rong Pan, "Test-cost sensitive classification on data with missing values," in IEEE Transactions on Knowledge and Data Engineering, Vol.18, No.5, pp.626-638, 2006.

[9] G. Madhu, "A Non-Parametric Discretization Based Imputation Algorithm for Continuous Attributes with Missing Data Values", International Journal of Information Processing, Volume 8, No.1, pp.64-72, 2014.

[10] Sreehari Rao, NareshKumar, "A New Intelligence-Based Approach for Computer-Aided Diagnosis of Dengue Fever, " , IEEE Transactions on Information Technology in Biomedicine, Vol.16, Issue 1, pp.112 – 118, 2012.

[11] V. Sree Hari Rao and M.NareshKumar, "Novel Approaches for Predicting Risk Factors of Atherosclerosis", IEEE Journal of Biomedical and Health Informatics, Vol.17,Issue1 pp. 183 – 189,2013

[12] G.Madhu, "A novel index measure imputation algorithm for missing data values: A machine learning approach", IEEE International Conference on Computational Intelligence & Computing Research, pp.1-7,2012.

[13] G.Madhu," A Novel Discretization Method for Continuous Attributes: A Machine Learning Approach", International Journal of Data Mining and Emerging Technologies, pp.34-43, Vol.4, No.1, 2014.

[14] G.Madhu," Improve the Classifier Accuracy for Continuous Attributes in Biomedical Datasets Using a New Discretization Method", Journal Procedia Computer Science,Vol 31, pp.671-79, 2014.

[15] Atif Khan, John A. Doucette, and Robin Cohen, " Validation of an ontological medical decision support system for patient treatment using a repository of patient data: Insights into the value of machine learning", ACM Trans. Intell. Syst. Technol,Vol.4,No.4,Article 68, 31 pages,2013.

[16] Jau-Huei Lin and Peter J. Haug,"Exploiting missing clinical data in Bayesian network modeling for predicting medical problems", Journal of Biomedical Informatics, Vol.41, Issue 1, pp.1-14, 2008.

[17] Karla L. Caballero Barajas and Ram Akella," Dynamically Modeling Patient's Health State from Electronic Medical Records: A Time Series Approach", In Proceedings of the 21th ACM SIGKDD International Conference on Knowledge Discovery and Data Mining (KDD'15), pp..69-78, 2015.

[18] Zhenxing Qin, Shichao Zhang, and Chengqi Zhang. 2006. Missing or absent? A Question in Cost-sensitive Decision Tree. In Proceedings of the 2006 conference on Advances in Intelligent IT: Active Media Technology, Yuefeng Li, Mark Looi, and Ning Zhong (Eds.). IOS Press, pp.118-125,2006.

[19] Shobeir Fakhraei, Hamid Soltanian-Zadeh, Farshad Fotouhi, and Kost Elisevich," Effect of classifiers in consensus feature ranking for biomedical datasets", In Proceedings of the ACM fourth international workshop on Data and text mining in biomedical informatics, DTMBIO '10,pp.67-68, 2010.

[20] Wei-Chao Lin, Shih-Wen Ke, Chih-Fong Tsai, "CANN: An intrusion detection system based on combining cluster centers and nearest neighbors", Knowledge-Based Systems, Volume 78, pp.13-21,2015.

[21] Aljawarneh, S., Shargabi, B., & Rashaideh, H. (2013). Gene classification: A review. Proceedings of IEEE ICIT.